\documentclass[prl,aps,amssymb,shownopacs,twocolumn,10pt,floatfix]{revtex4-1}
\usepackage{amsmath}
\usepackage{amssymb}
\usepackage{amsthm}
\usepackage{amsfonts}
\usepackage{listings}
\usepackage{enumerate}
\usepackage{latexsym}
\usepackage{hyperref}

\usepackage{psfrag}
\usepackage{color}
\usepackage{bm}
\usepackage[pdftex]{graphicx}
\usepackage{subfigure}

\def\ket#1{\left|#1 \right\rangle}

\def\braket#1#2{\left\langle #1 | #2 \right\rangle}
\def\matrix22#1#2#3#4{\left(\begin{array}{cc}#1&#2\\#3&#4\end{array}\right)}

\usepackage{ulem}

\begin{document}

\title{Bosonic integer quantum Hall effect in optical flux lattices}

\author{A. Sterdyniak$^1$}
\author{Nigel R. Cooper$^{2}$}
\author{N. Regnault$^{3,4}$}
\affiliation{$^1$ Institute for Theoretical Physics, University of Innsbruck, A-6020 Innsbruck, Austria\\
$^2$ T.C.M. Group, Cavendish Laboratory, J.J. Thomson Avenue, Cambridge CB3 0HE, United Kingdom\\
$^3$ Department of Physics, Princeton University, Princeton, NJ 08544\\
$^4$ Laboratoire Pierre Aigrain, Ecole Normale Sup\'erieure-PSL Research
University, CNRS, Universit\'e Pierre et Marie Curie-Sorbonne Universit\'es,
Universit\'e Paris Diderot-Sorbonne Paris Cit\'e, 24 rue Lhomond, 75231
Paris Cedex 05, France}

\begin{abstract}
In two dimensions strongly interacting bosons in a magnetic field can realize a bosonic integer quantum Hall state, the simplest two dimensional example of a symmetry protected topological phase. We propose a realistic implementation of this phase using an optical flux lattice. Through exact diagonalization calculations, we show that the system exhibits a clear bulk gap and the topological signature of the bosonic integer quantum Hall state. In particular, the calculation of the many-body Chern number leads to a quantized Hall conductance in agreement with the analytical predictions. We also study the stability of the phase with respect to some of the experimentally relevant parameters. 
\end{abstract}

\date{\today}
\maketitle

\paragraph{Introduction ---}

Symmetry Protected Topological (SPT) insulating phases of interacting bosons are the analogues of the celebrated free fermion topological insulators for bosonic systems (see Ref.~\onlinecite{Senthil-annurev-conmatphys-031214-014740} for a short review). The SPT phases have a bulk gap but no intrinsic topological order. As a consequence, they do not host any fractionalized excitations. Still they can exhibit protected boundary excitations. Non-interacting boson phases are topologically trivial. The need for strong interactions to obtain SPT phases makes them harder to study than their fermionic cousins, where the band topology is the only required information. Even their cohomology classification\cite{Fidkowski-PhysRevB.83.075103,Turner-PhysRevB.83.075102,SchuchPhysRevB.84.165139,Chen21122012,Chen-PhysRevB.87.155114} does not cover all the known examples\cite{Vishwanath-PhysRevX.3.011016,Wang-PhysRevB.87.235122,Metlitski-PhysRevB.88.035131,Burnell-PhysRevB.90.245122,Kapustin-2014arXiv1403.1467K,
Kapustin-2014arXiv1404.6659K}. Therefore, finding experimentally relevant microscopic models in dimension larger than one remains an important and difficult task.
 
The prototype of a SPT phase in two dimensions is the boson integer quantum Hall (bIQH) state. The physical properties of this state have been studied in Refs.~\onlinecite{Lu-PhysRevB.86.125119,Senthil-PhysRevLett.110.046801,Liu-PhysRevLett.110.067205,Geraedts2013288}. In particular, the Hall conductivity was shown\cite{Lu-PhysRevB.86.125119} to be quantized and equal to an even integer.  The edge physics consists of a charged chiral edge mode and a counter propagating neutral mode. Despite being non-chiral this edge structure is protected so long as charge conservation symmetry is preserved. Following the proposal of Ref.~\onlinecite{Senthil-PhysRevLett.110.046801}, several recent numerical studies\cite{Furukawa-PhysRevLett.111.090401,Wu-PhysRevB.87.245123,Regnault-PhysRevB.88.161106} have a pointed out the possible emergence of bIQH in bilayer bosonic fractional Hall system at filling factor $\nu=1$ for each component. In particular, a robust gapped phase has been observed, and some indications 
of 
the 
topological nature has been obtained. While providing a proof of concept, this setup would be difficult to realize in an experimental ultracold atomic system. Even the numerical simulations still lack direct evidence of the topological nature of the observed state, such as the existence of edge modes or the measurement of the Hall conductance.

Motivated by experimental settings in ultracold atomic gases, lattice versions with ``Chern bands'' could replace the lowest Landau level for the bIQH while providing a realistic implementation. In such systems, the emergence of strongly correlated topological phases has been studied in both the context of the fractional quantum Hall (FQH) effect in the presence of a lattice\cite{Kol:1993p82,Srensen:2005p58,Palmer:2006p63,Hafezi:2007p67,Palmer-PhysRevA.78.013609,Moller:2009p184,Moller:2012p2646,Sterdyniak-PhysRevB.86.165314,Scaffidi-PhysRevB.90.115132} or in a fractionally filled Chern band, namely a fractional Chern insulator\cite{neupert-PhysRevLett.106.236804,sheng-natcommun.2.389,regnault-PhysRevX.1.021014,BERGHOLTZ-JModPhysB2013,Parameswaran2013816}. As opposed to a single Landau level, lattice models can be engineered to have bands with a Chern number $C$ higher than one. Since in the non-interacting regime these bands can be viewed as $C$ copies of a Chern one band, a single $C=2$ band would mimic a 
bilayer system. 
The fate of a partially filled band carrying a Chern number $C>1$ in the strongly interacting regime has been recently studied\cite{Barkeshli-PhysRevX.2.031013,Wang-PhysRevB.86.201101,Yang-PhysRevB.86.241112,Liu-PhysRevLett.109.186805,Sterdyniak-PhysRevB.87.205137,Wu-PhysRevLett.110.106802,Wu-PhysRevB.91.041119,Moeller-2015arXiv150406623M}. Some of these systems host new phases that are generalizations of the Halperin states~\cite{Halperin83} in the FQH with color-orbit couplings. 
It has also been shown that the $C=2$ band of the Harper-Hofstadter model can support a non-fractionalized bosonic phase\cite{Moller:2009p184,Moeller-2015arXiv150406623M}  at least at low particle densities. Still, a simple lattice model of a Chern insulator in which the bIQH state can appear at high particle densities remains lacking.

In this letter, we consider the implementation of the bIQH in optical flux lattices~\cite{Cooper-PhysRevLett.106.175301,cooper-EuroPhyLett2011,Cooper-PhysRevLett.109.215302,Sterdyniak-PhysRevB.91.035115} as a potential experimental realization in ultracold atomic gases. Optical flux lattices can be designed in several ways, offering a large control over the band topology, dispersion and Berry curvature distribution through parameters such as the number of internal states and the laser couplings~\cite{Cooper-PhysRevLett.109.215302}. These systems also allow to tune the Chern number of the lowest band while preserving its approximate flatness. In presence of strong interaction, we have shown the emergence of Halperin-like states\cite{Sterdyniak-PhysRevB.91.035115}. Thus they constitute a natural candidate to look for the bIQH. Using exact diagonalization, we provide compelling evidence for the emergence of such a phase when fully filling the lowest band of a lattice for which the Chern number has been set to 
two. In particular, we compute the many-body Chern number, thus showing the quantization of the Hall conductivity.

We consider bosonic atoms with $N$ internal degrees of freedom in the optical flux lattice model introduced by Ref.~\onlinecite{Cooper-PhysRevLett.109.215302}. The one-body Hamiltonian is given by
\begin{equation}
\hat{H} = \frac{{\bm P}^2}{2M} \hat{\openone}_N + \hat{V}({\bm r})
\label{eq::OFL_Ham}
\end{equation}
where $\hat{\openone}_N$ is the $N \times N$ identity matrix. The coupling of the different internal degrees of freedom using two-photon Raman transitions is described by the potential $\hat{V}({\bm r})$, with a characteristic scale $V$ set by the laser strength. The laser beams are arranged to form a triangular lattice, and induce a set of allowed momentum transfers in reciprocal space with a triangular pattern. The characteristic kinetic energy scale in the Hamiltonian is the recoil energy $E_R= \frac{\hbar^2\kappa^2}{2M}$ where $\kappa$ is the wave number of the laser beams.

\begin{figure}[htb]
\includegraphics[width=0.9\columnwidth]{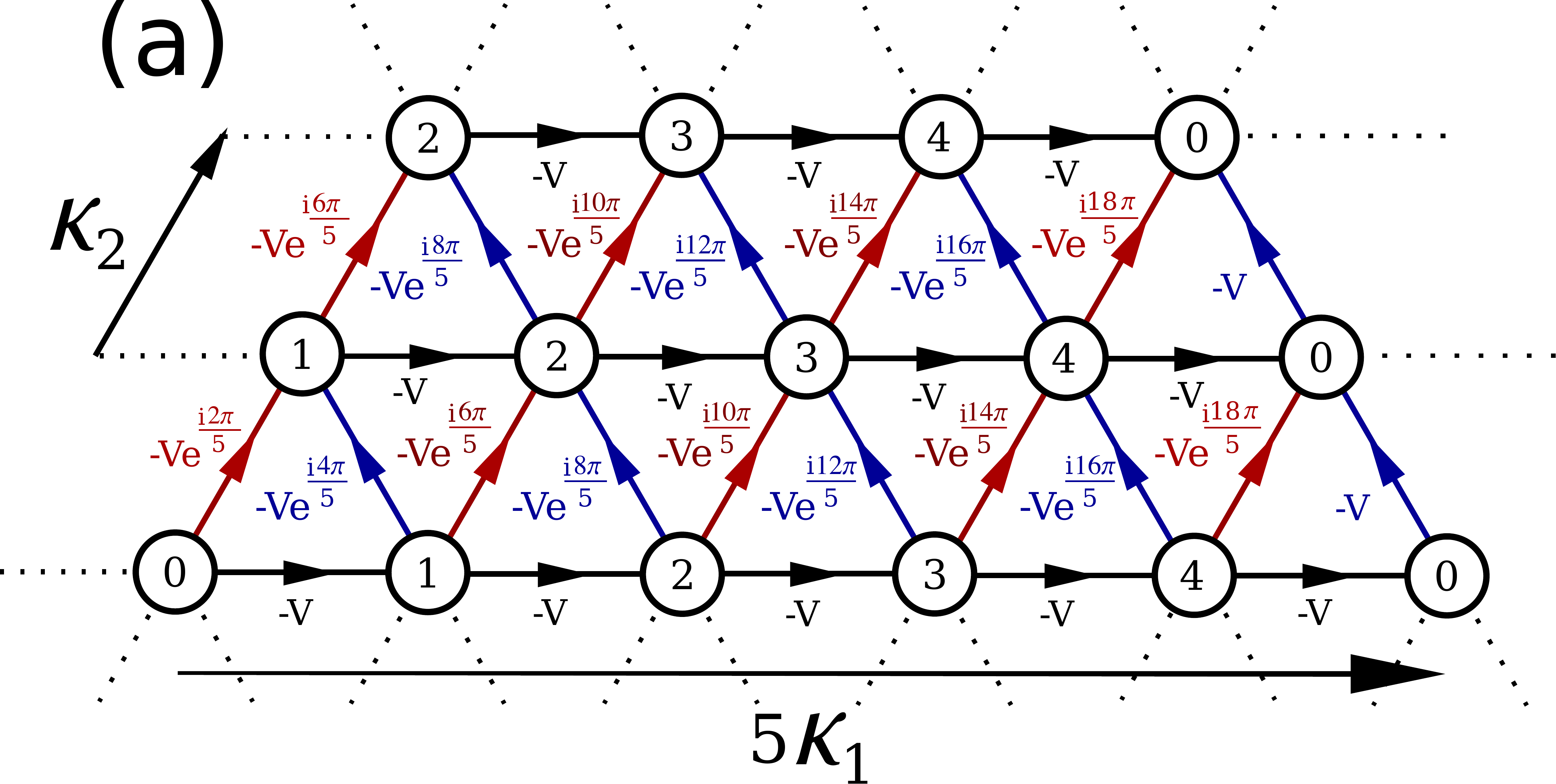}
\includegraphics[width=0.9\columnwidth]{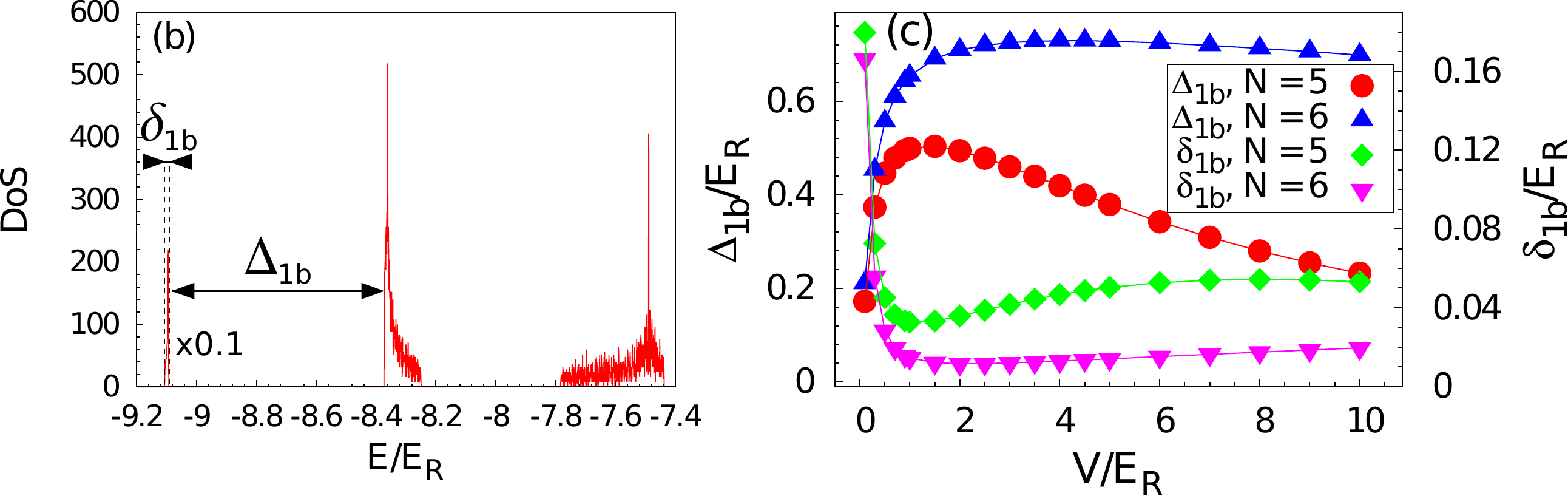}
\caption{(a) Schematic description of the first Brillouin zone of the OFL model that we have considered for $N=5$. The blue, red and black links represents the hopping amplitudes defined by the potential $\hat{V}({\bm r})$. The phases  are set to produce a lowest band carrying a Chern number $C=2$. (b) Density of states for $V/E_R=3$ and $N=6$. Note that we have rescaled the density of states of the lowest band by a factor $0.1$. The band spread of the lowest band is $\delta_{1{\rm b}} = 0.0126 E_R$ and the band gap between the lowest band and the second band is $\Delta_{1{\rm b}} =0.721 E_R $. (c) Band spread and band gap of the lowest band as a function of the laser strength $V$ for $N=5$ and $N=6$.}
\label{fig::ofl}
\end{figure}

A synthetic gauge field in reciprocal space can be created by controlling the relative phases between the different laser beams. These phases can be chosen so that the lowest band possesses any Chern number $C$ smaller than $N$, the ratio $C/N$ controlling the band dispersion. In the following, we will focus on $C=2$. More details on the one-body model can be found in Ref.~\cite{Cooper-PhysRevLett.109.215302,Sterdyniak-PhysRevB.91.035115}. In Fig.~\ref{fig::ofl}a, we give a schematic description of the lattice in reciprocal space. It is a triangular lattice spanned by the vectors ${\bm  \kappa}_1= (1,0)\kappa$ and ${\bm \kappa}_2=\left(\frac{1}{2},\frac{\sqrt{3}}{2}\right)\kappa$. Due to a higher translation symmetry the first Brillouin zone is spanned by $N{\bm  \kappa}_1$ and ${\bm  \kappa}_2$. As a result, the real space unit cell spanned by ${\bm a}_1=\frac{2\pi}{N\kappa}(1,-1/\sqrt{3})$ and ${\bm a}_2=\frac{4\pi}{\kappa\sqrt{3}}(0,1)$, has an aspect 
ratio equal to $N$. A typical density of states is shown in Fig.~\ref{fig::ofl}b. It clearly shows the flatness of the lowest band, quantified by its spread $\delta_{1{\rm b}}$, and the large gap $\Delta_{1{\rm b}}$ to the second band. In this letter, we will consider the two numbers of internal degrees of freedom $N=5$ and $N=6$. At the level of the one-body spectrum, we already observe in Fig.~\ref{fig::ofl}c a sharp difference between these two cases with respect to the laser strength. Indeed, the gap is rapidly decreasing with $V$ for $N=5$ while it stays mostly unchanged for $N=6$.

While Chern insulators are defined by a tight-binding model, optical flux lattice models are continuous in real space. Thus, the interaction between the atoms mainly depends on the considered atomic species. Here, we focus on the simplest and more realistic interaction: the $s$-wave scattering that correctly describes cold gases of alkali atoms like $^{87}$Rb. Thus, the interaction potential is given by 
\begin{equation}
{\cal H}_{\rm int} = V_{\rm int}\delta({\bm r}-{\bm r}') \,.
\label{eq::int_delta}
\end{equation}
Previous studies\cite{Furukawa-PhysRevLett.111.090401,Wu-PhysRevB.87.245123,Regnault-PhysRevB.88.161106} on the emergence of the bIQH within a bosonic bilayer FQH system involved interactions both within each layer and between the two layers, also given by the interaction potential of Eq.~\ref{eq::int_delta}. A major difference in that case is the ability to tune the ratio of the interaction strengths between the layers and within a layer. Indeed, the bIQH was found\cite{Regnault-PhysRevB.88.161106} for ratios between 0.8 and 1.3.  While a Chern $C=2$ band can be decoupled into two copies of a Chern one band, this cannot be achieved once interactions are included. As a consequence, the notion of intra and inter layer interaction is not meaningful in this context and the interaction strength ratio is locked to one\cite{Wu-PhysRevLett.110.106802}. 

We now turn to the numerical study of the model. We consider $N_B$ bosons on a finite size system with periodic boundary conditions defined by the vectors ${\bm L_x} = N_x {\bm a}_1$ and ${\bm L_y} = N_y {\bm a}_2$. The number of orbitals per band is then equal to $N_xN_y$ and, thus, the filling factor is $\nu=\frac{N_B}{N_xN_y}$. Here, we focus on $\nu$ equal or close to 1, i.e. one boson per state of the lowest $C=2$ band.

Similar to the lowest Landau level projection in the numerical studies of FQH effect, we project this interaction onto the lowest band of the one-body model, removing the effect of both band mixing and band dispersion. Such an approximation is valid as long as $\delta_{1{\rm b}}\ll V_{\rm int} \ll \Delta_{1{\rm b}}$. At unit filling of fermions, this approximation would  lead to the usual integer quantum Hall (IQH) effect, the interaction only shifting the system energy. This is not case for the bosonic atoms considered where the non-interacting and strong interacting regimes differ.

\begin{figure}[htb]
\includegraphics[width=0.49\columnwidth]{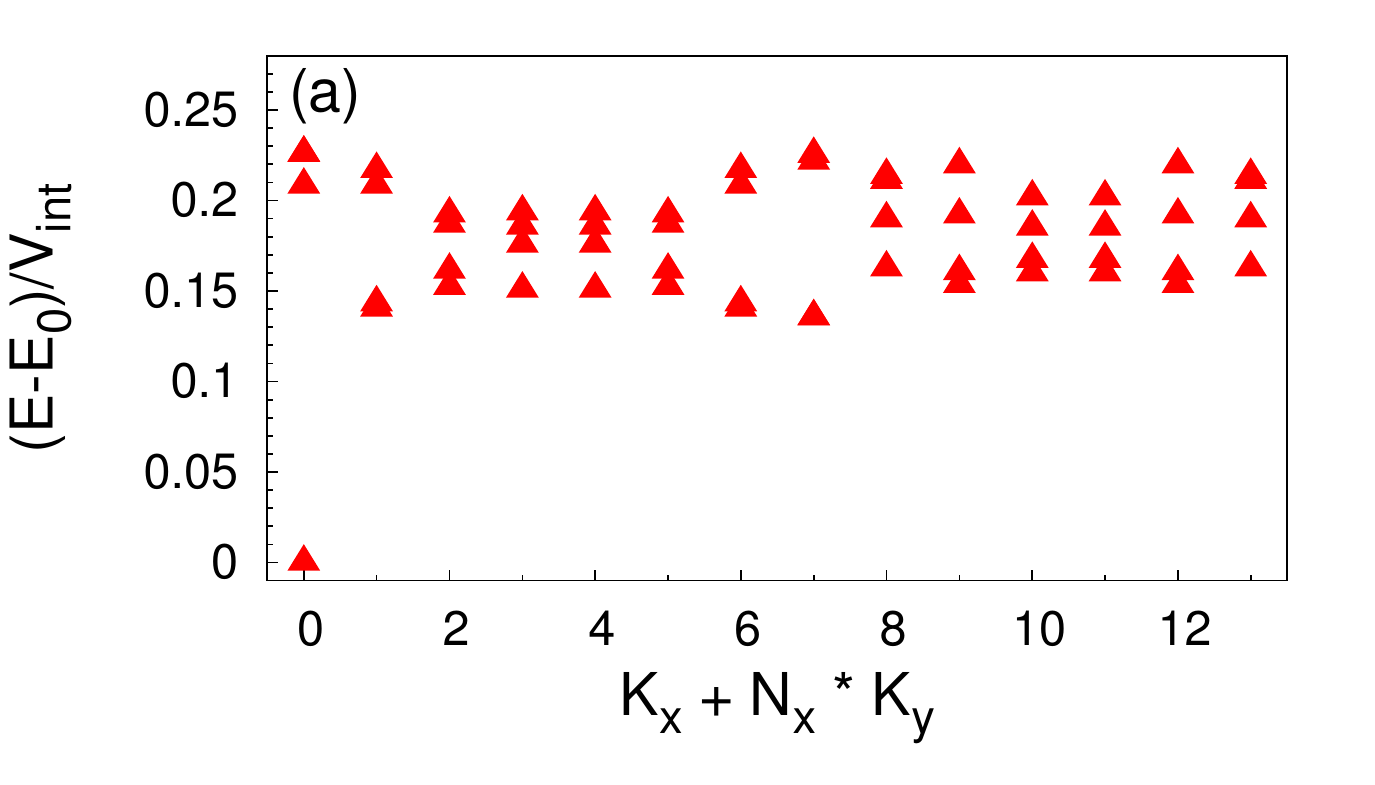}
\includegraphics[width=0.49\columnwidth]{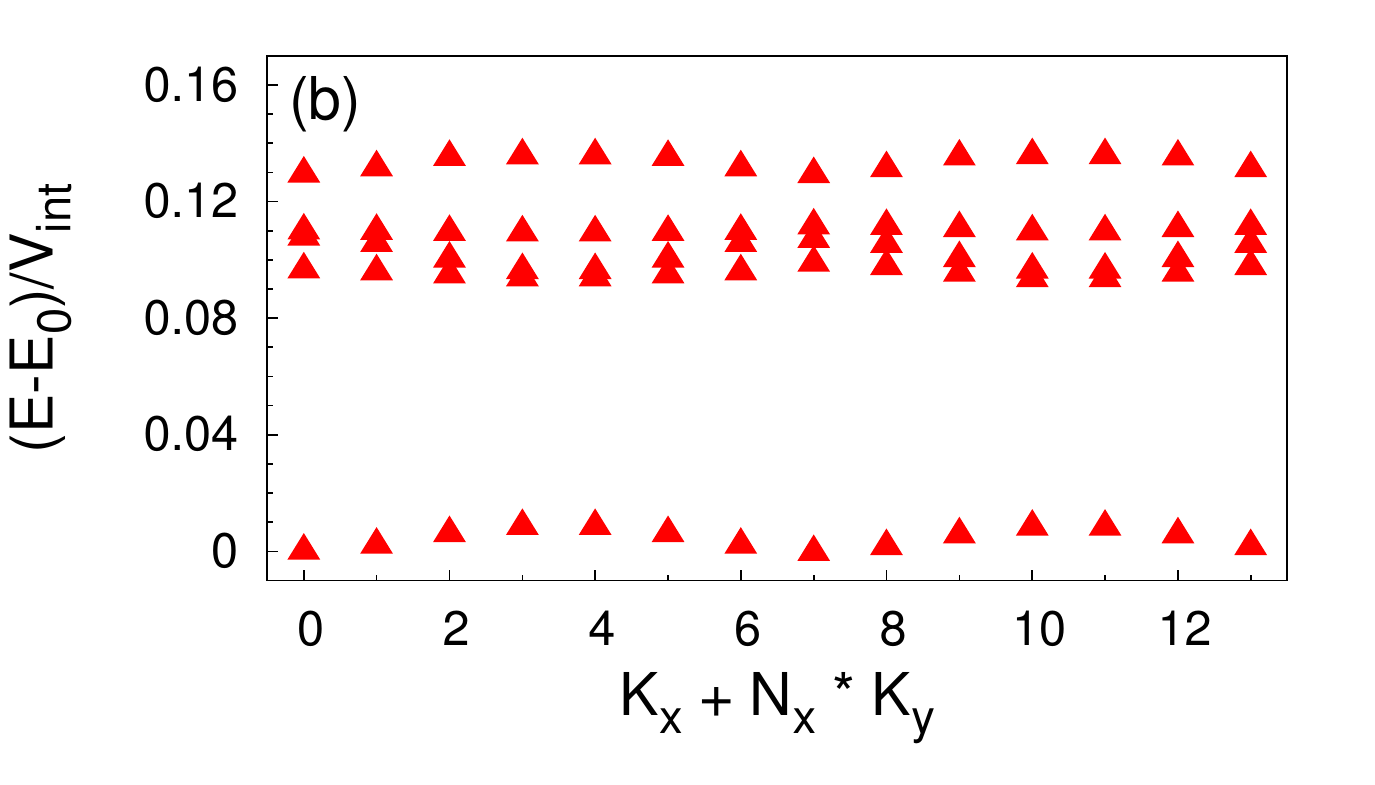}
\caption{\textit{Left Panel:} Low energy spectrum of the two body interaction at $\nu=1$ for $N=6$, $N_B=14$, $N_x=7$, $N_y=3$ and $V=3E_R$. The energies are plotted as a function of the linearized momentum index where $(K_x,K_y)$ are the two integers defining the momentum sector. Note that the energies are shifted by the ground state energy $E_0$. \textit{Right Panel:} Energy spectrum of the two body interaction at $\nu=1$ for $N=6$, $N_B=13$, $N_x=5$, $N_y=3$ and $V=3E_R$. Compared to the left panel, there is one particle less and thus one added quasihole. We observe an almost flat band of low energy states with one state per sector as expected for the IQH effect.}
\label{fig::figIFQH}
\end{figure}

A typical energy spectrum at unit filling is shown in Fig.~\ref{fig::figIFQH}a. It exhibits a single ground state with a zero momentum separated from the excited states by an energy gap $\Delta$. This is expected for the bIQH : having no intrinsic topological order, its degeneracy should be equal to one irrespective of the genus of the surface we consider, including our torus geometry. Other phases that have been proposed in such a system, like two copies of the Moore-Read state\cite{Moore:1991p165} or the non-abelian spin singlet state\cite{Ardonne-PhysRevLett.82.5096}, would exhibit a non-trivial degeneracy. Note that we observe this single state at zero momentum separated by an energy gap irrespective of the particle number, showing the absence parity effect: in a bilayer setup, $N_B$ is required to be even by construction.

Another feature that can be tested is the number of quasiholes states. For Chern insulators, this can be accessed by increasing the number of sites compared to the ground state at fixed number of particles or by removing particles at fixed number of sites. Here, we choose the latter solution and show in Fig.~\ref{fig::figIFQH}b, the energy spectrum with one particle less than in Fig.~\ref{fig::figIFQH}a. This spectrum exhibits an almost flat lowest energy band consisting of one level per momentum sector. This is analogous to the IQHE where the number of states when adding a quasihole is equal to the number of orbitals in each Landau levels.

\begin{figure}[htb]
\includegraphics[width=0.9\columnwidth]{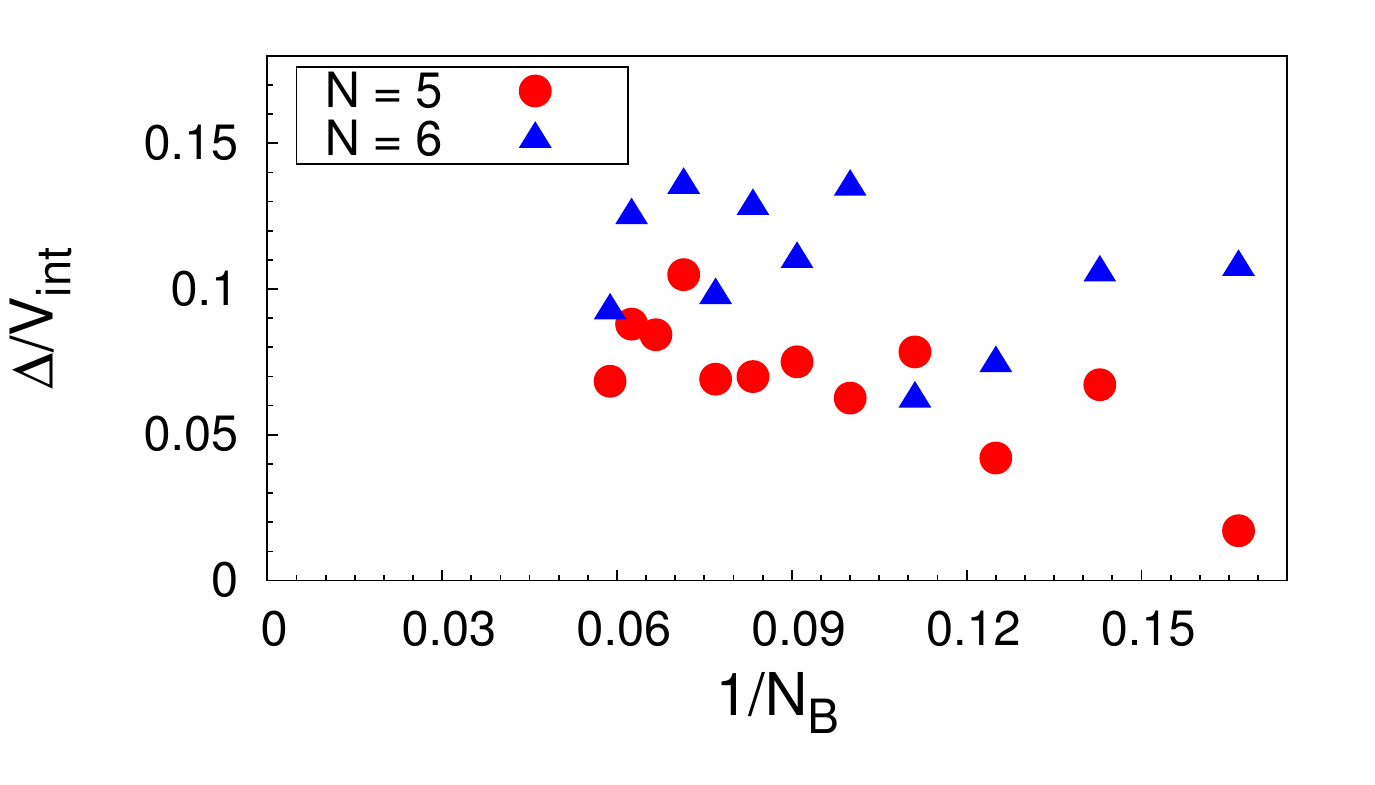}
\caption{Gap as a function of the inverse of the particle number for $N=5$ (red dots) and $N=6$ (blue triangles) and $V=3E_R$ up to $N_B=17$. For each particle number, we choose $N_x$ and $N_y$ so that the aspect ratio is the closest possible to $1$. However, there is still important aspect ratio variations between the different point which prevent from doing an relevant extrapolation of the gap at the thermodynamical limit. Quite clearly, this gap does not close at the thermodynamical limit and will be of order $0.1V_{\rm int}$}
\label{fig::gap_extropolation}
\end{figure}

To show that the features of the bIQH persist in this system in the thermodynamic limit, we show the evolution of the many-body gap $\Delta$ as a function of system size in Fig.~\ref{fig::gap_extropolation} for $N=5$ and $N=6$. While the gap exhibits some fluctuations resulting from the aspect ratio variation (a known effect\cite{Wu-PhysRevB.85.075116,PhysRevLett.111.126802} for small sizes in any fractional Chern insulator), it shows a tendency to converge for larger systems around $\Delta\simeq 0.1 V_{\rm int}$. Note that we have also considered lower values of $N$ such as $N=4$. There size effects are even more pronounced, including the absence of a gap in some cases. 

\begin{figure}[htb]
\includegraphics[width=0.9\columnwidth]{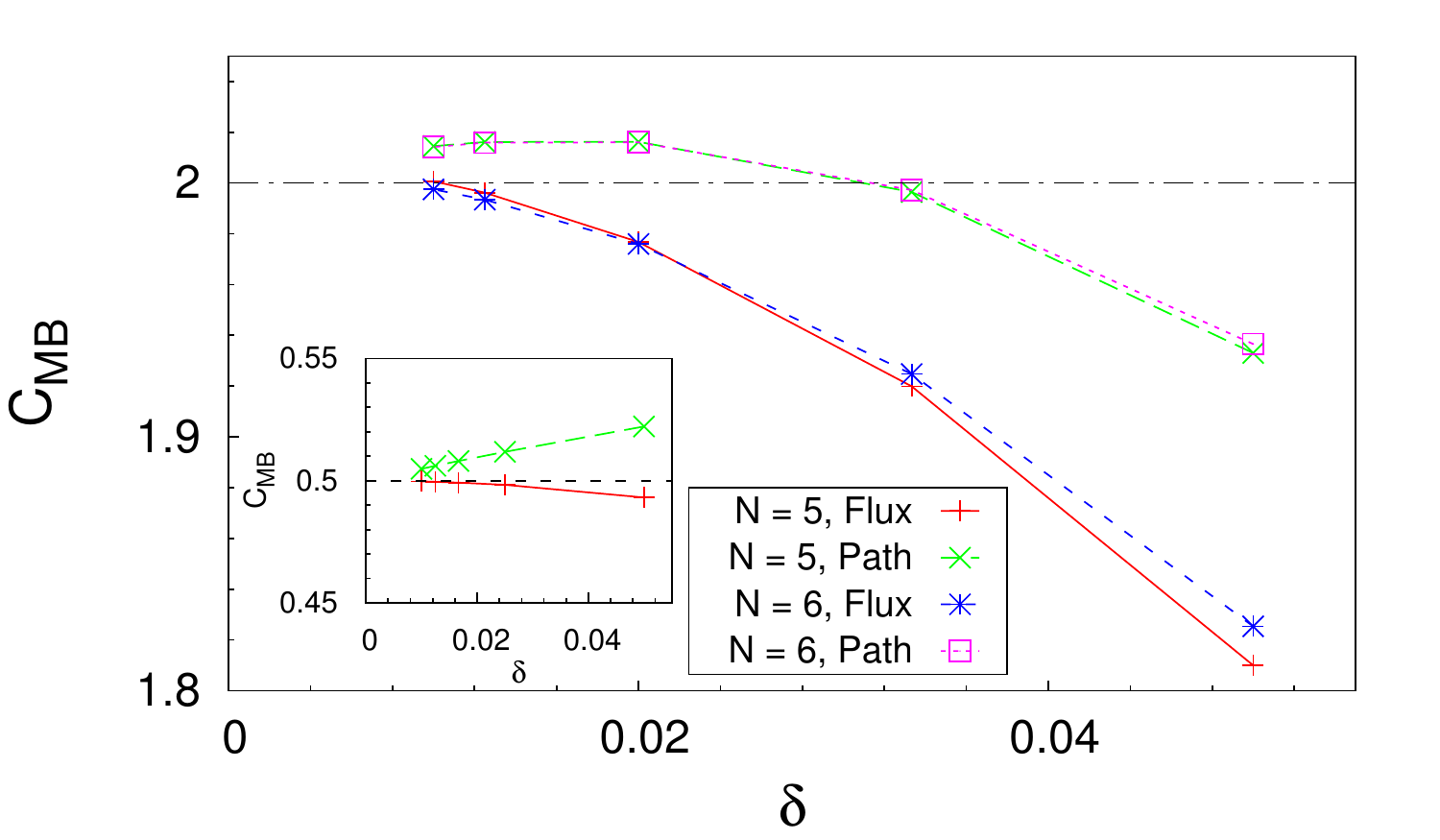}
\caption{Many-body Chern number $C_{\rm MB}$ computed both using the surface integral and the path integral version, as function of the discretization step $\delta$ of the $(\theta_x,\theta_y)$ torus for $N=5$ and $N=6$. The system considered here has $N_B=12$, $N_x=6$,$N_y=2$ and $V=3E_R$. The dashed horizontal line at $C_{\rm MB}=2$ is a guide for the eye. In the inset, we show the result at $\nu=\frac{1}{2}$ when the lowest band has $C=1$ where Laughlin state is realized for $N_B=6$, $N_x=4$, $N_y=3$ and $N=5$.}
\label{fig::Chern_number}
\end{figure}

To establish on a stronger footing the realization of the bIQH in this system, we compute the many-body Chern number of the ground state $\ket{\Psi}$. A non-zero value of this number would show the quantization of the Hall conductance, a direct signature of the bIQH. The many-body Chern number can be computed either via the contour integal
\begin{equation}
C_{\rm MB} = \frac{1}{2\pi}\oint \textrm{Im} \braket{\Psi}{\nabla\Psi}
\end{equation}
or, equivalently, using Stokes' theorem, by the surface integral: 
\begin{eqnarray}
C_{\rm MB} =\frac{1}{2\pi}\int\int d\theta_xd\theta_y \textrm{Im} \left(\braket{\frac{\partial \Psi}{\partial \theta_x}}{\frac{\partial \Psi}{\partial \theta_y}} -  \braket{\frac{\partial \Psi}{\partial \theta_y}}{\frac{\partial \Psi}{\partial \theta_x}}\right)\nonumber\\
\end{eqnarray}
where $\theta_x$ and $\theta_y$ are the two fluxes that can be inserted through the two independent non-contractible loops on the torus. While these two integrals are equal, their numerical evaluation requires dividing the ($\theta_x$,$\theta_y$) torus in small patches of size $2\pi\delta \times 2\pi\delta$. This discretization yields slightly different results. We show the many-body Chern number computed using both methods for a system of $N_B=12$ bosons with $N=5$ and $N=6$ as a function of $\delta$ in Fig.~\ref{fig::Chern_number}. Generically, the contour integral tends to overestimate the result whereas the surface one underestimates it. We have checked our numerical procedure against the half filled $C=1$ band where Laughlin $\nu=1/2$ physics is realized~\cite{Sterdyniak-PhysRevB.91.035115} and we have found a similar behavior. These results are shown in the inset of Fig.~\ref{fig::Chern_number}. In both cases, we found that the Chern number is equal to $2$ within $1\%$ precision. We have obtained similar 
results for all systems from $N_B=9$ to $N_B=12$, irrespective of the parity of $N_B$. This is an unambiguous evidence of the Hall conductance quantization and thus the realization of the bosonic integer quantum Hall effect in this system.
 
\begin{figure}[htb]
\includegraphics[width=0.75\columnwidth]{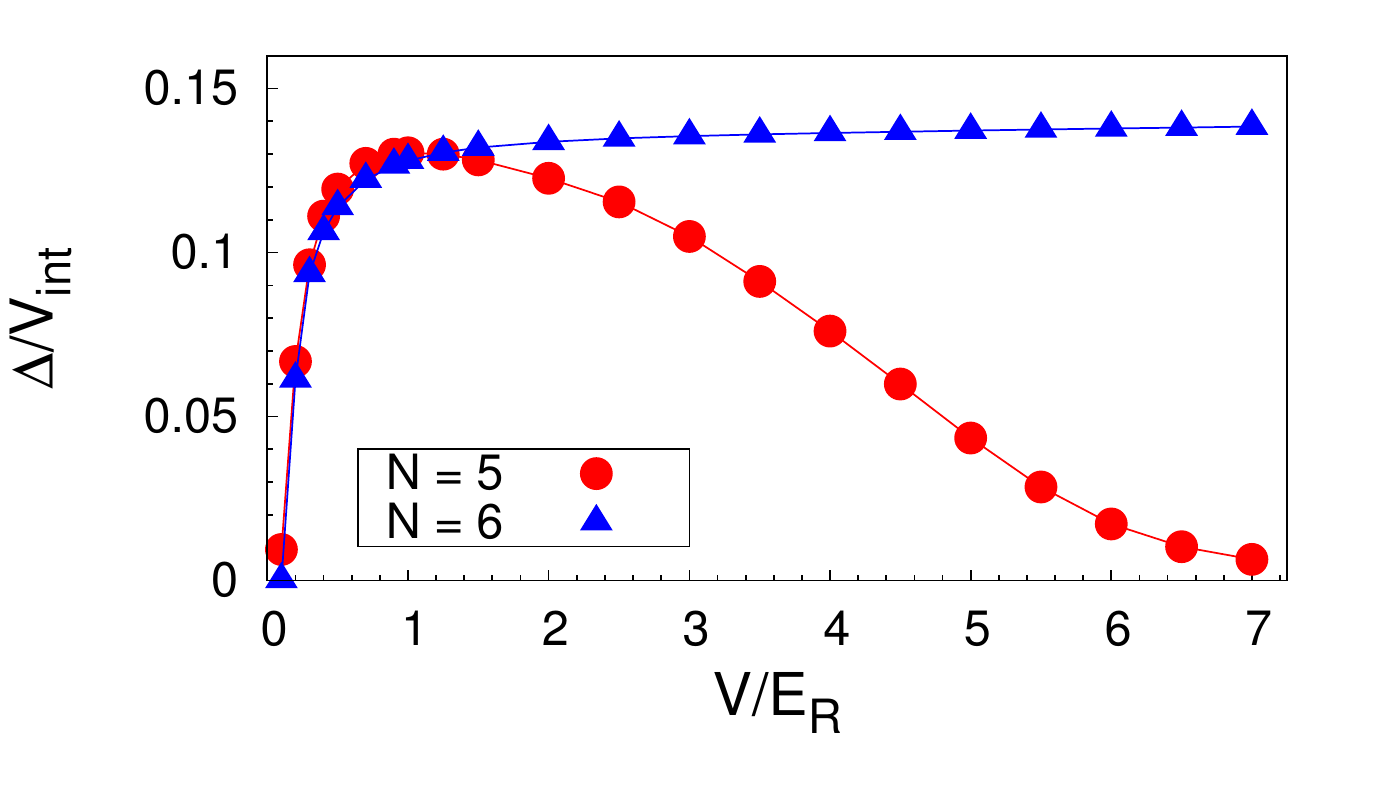}
\caption{Influence of the laser strength $V$ on the many-body gap $\Delta$ for $N_B=14$, $N_x=7$ and $N_y=2$. We have considered both $N=5$ (red dots) and $N=6$ (blue triangles). The behavior of $\Delta$ exactly mimics the behavior of the one-body gap $\Delta_{1{\rm b}}$ shown in Fig.~\ref{fig::ofl}c.}
\label{fig::gap_vs_las}
\end{figure}

\begin{figure}[htb]
\includegraphics[width=0.85\columnwidth]{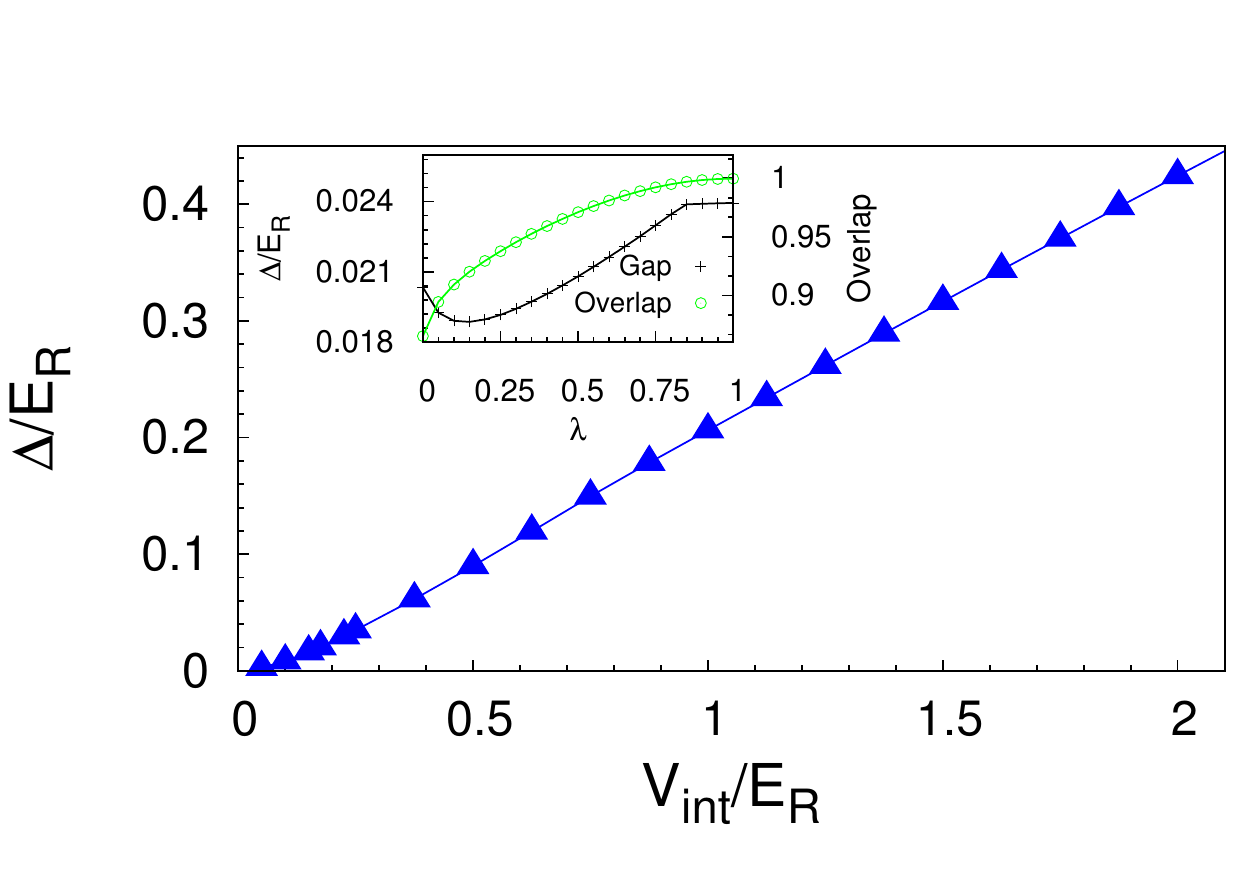}
\caption{Many-body gap as a function of the interaction strength $V_{\rm int}$ when the two lowest bands and their dispersion relations are taken into account, for $N=6$, $N_B=10$, $N_x=10$, $N_y=1$ and $V=3E_R$. For comparison, the one-body gap is $\Delta_{1{\rm b}} =0.721 E_R $. The inset gives the continuous evolution at $V_{\rm int}=0.175 E_R$ from the two band model ($\lambda=0$) to the flat band limit ($\lambda=1$). Both evolutions of the gap (black line) and the overlap with the flat band ground state (green line) show the adiabatic connection between these two systems.}
\label{fig::gap_vint}
\end{figure}

Until now, we have focused on a specific set of parameters for our model. It is experimentally relevant to look at the effect of the laser strength or the flat band approximation on the bIQH phase stability. We first consider the effect of the laser strength in the flat band approximation. The numerical results for $N_B=14$ are shown in Fig.~\ref{fig::gap_vs_las}. Remarkably, we observe that the many-body gap has the same behavior as the one-body gap $\Delta_{1{\rm b}}$ that we have computed in Fig.~\ref{fig::ofl}c. The maximum for $N=6$ is in the regime $V\simeq E_R$ of a shallow optical lattice, emphasising that this system is far from the tight-binding limit. A more important question is the fate of the bIQH phase away from the flat band approximation. Due to the numerical limitation, we still have to truncate the one-body basis. We thus consider the two lowest bands including their dispersion relation and look at the effect of the interaction strength $V_{\rm int}$. The many-body gap is shown in Fig.~\ref{fig::gap_vint}.  We have not directly computed the Chern number in this case. However,  as interaction strength $V_{\rm int}$ increases and band mixing becomes more relevant, we find smooth evolution, with no gap collapse or phase transition. Thus, these states are adiabatically connected to the states formed in the projected lowest band, so must retain the same Chern number $C=2$ that we computed in that case. We have also checked that we can continuously deform the system up to the flat band limit without gap closing, as shown in the inset of Fig.~\ref{fig::gap_vint}.

\paragraph{Conclusion ---} In this letter, we have considered the realization of the $C=2$ bosonic integer quantum Hall effect in the lowest band of an optical flux lattice. We have characterized the topological order of the phase through the quashihole excitations and the evaluation of the electrical Hall conductance via many-body Chern number.\\

\emph{Note:} During the final stage of this letter, we became aware of a related paper Ref.~\cite{He-2015arXiv150601645H} providing numerical evidence for the bIQH in another lattice model with correlated hopping and a background gauge field.\\

\begin{center}ACKNOWLEDGEMENTS\end{center}

We acknowledge T. Senthil, G. M\"oller, B.A. Bernevig and F. Pollmann for useful discussions. A.S. acknowledges support through the Austrian Science Foundation (FWF) SFB Focus (F40-18). This work was supported by the Austrian Ministry of Science BMWF as part of the Konjunkturpaket II of the Focal Point Scientific Computing at the University of Innsbruck. N.R. was supported by Keck grant, ANR-12-BS04-0002-02 and the Princeton Global Scholarship.

\bibliography{biqhe_ofl}

\end{document}